\documentclass{article}
\usepackage[nonatbib, preprint]{neurips_2023}
\usepackage[utf8]{inputenc} 
\usepackage[T1]{fontenc}    
\usepackage[hidelinks]{hyperref}       
\usepackage{url}            
\usepackage{booktabs}       
\usepackage{amsfonts}       
\usepackage{amsmath}
\usepackage{amssymb}
\usepackage{amsthm}
\usepackage{mathtools}
\usepackage{nicefrac}       
\usepackage{microtype}      
\usepackage{xcolor}
\definecolor{darkgreen}{RGB}{0,100,0}
\definecolor{lavender}{RGB}{230, 230, 250}
\usepackage{dirtytalk}
\usepackage{lipsum}
\usepackage{subcaption}
\usepackage{graphicx}
\usepackage{cleveref}

\usepackage[
backend=biber,
style=nature,
sorting=none
]{biblatex}
\theoremstyle{definition}
\newtheorem{theorem}{Theorem}

\newtheorem{corollary}{Corollary}
\newtheorem{remark}{Remark}

\title{Bounding data reconstruction attacks with the hypothesis testing interpretation of differential privacy}

\author{
Georgios Kaissis\thanks{\texttt{\{g.kaissis, alex.ziller, daniel.rueckert\}@tum.de}}, 
Jamie Hayes\thanks{\texttt{jamhay@google.com}}, 
Alexander Ziller\footnotemark[1], 
Daniel Rueckert\footnotemark[1] \\ \\
Technical University of Munich\footnotemark[1], \; Google DeepMind\footnotemark[2]}

\begin{document}

\maketitle

\begin{abstract}
    We explore Reconstruction Robustness (ReRo), which was recently proposed as an upper bound on the success of data reconstruction attacks against machine learning models.
    Previous research has demonstrated that differential privacy (DP) mechanisms also provide ReRo, but so far, only asymptotic Monte Carlo estimates of a tight ReRo bound have been shown. 
    Directly computable ReRo bounds for general DP mechanisms are thus desirable. 
    In this work, we establish a connection between hypothesis testing DP and ReRo and derive closed-form, analytic or numerical ReRo bounds for the Laplace and Gaussian mechanisms and their subsampled variants. 
\end{abstract}

\section{Introduction}
In the rapidly advancing field of machine learning (ML), the importance of preserving privacy cannot be understated, particularly in critical tasks where privacy may be compromised through attacks on unprotected ML models.
Among these, membership inference (MI) poses a considerable risk \cite{carlini2022membership}.
Here, an adversary attempts to determine whether a candidate record was part of the model's training database. 
Differential privacy (DP) \cite{dwork2006differential} plays a crucial role as a safeguard against privacy risks in ML. 
Its guarantees can be interpreted in terms of the protection it offers against MI, a notion termed the \textit{hypothesis testing interpretation of DP} \cite{dong2021gaussian}.
Broadly speaking, protecting against MI also serves to protect against all weaker forms of attack \cite{balle2022reconstructing}.
For example, data reconstruction (DR) attacks \cite{carlini2021extracting, carlini2023extracting}, where adversaries attempt to extract records from the model's weights or gradients \cite{geiping2020inverting}, are also prevented by DP mechanisms.
In fact, it can be shown that protecting against DR requires substantially less noise than protecting against MI \cite{balle2022reconstructing}.
Recent works have proposed formal bounds tailored to DR.
For instance, Guo et al. \cite{guo2022bounding} frame DR as a signal estimation problem and use the properties of the Fisher information matrix to lower-bound reconstruction error.
Moreover, Guo et al. \cite{guo2022analyzing} utilise Fano's inequality to bound the mutual information between the training data and the model's parameters.
Last but not least, Balle et al. \cite{balle2022reconstructing} recently proposed Reconstruction Robustness (ReRo), which serves as a high-probability bound on successful DR.
Moreover, this work's authors prove a strong relationship between DP and ReRo in the sense that (Rényi-)DP \cite{mironov2017renyi} implies ReRo (and \textit{vice versa} under some preconditions).
Very recently, Hayes et al. \cite{hayes2023bounding} strengthened the aforementioned results by circumventing the utilisation of Rényi-DP and bounding ReRo directly.

In this work, we expand upon the previous investigations on ReRo, which we regard as the most promising DR bound (as it both outperforms previous DR guarantees and is closely matched by the results of empirical DR attacks against ML models).
The aforementioned work by Hayes et al. \cite{hayes2023bounding} limits its purview to DP-SGD \cite{abadi2016deep} and utilises a Monte Carlo (MC) technique to estimate the ReRo bound. 
This MC bound only holds asymptotically and cannot be used efficiently in workflows involving large datasets.
Methods to directly obtain ReRo upper bounds for arbitrary datasets and mechanisms (e.g. also the Laplace mechanism and its subsampled variant), would thus be of value to practitioners. 

\paragraph{Contributions}
The contributions of our work are as follows:
(1) We extend the work of Hayes et al. by proposing ReRo bounds derived from the hypothesis testing interpretation of DP.
(2) We furnish closed-form bounds for the Gaussian and Laplace mechanisms and provide an analytic formulation for the Poisson-sampled Gaussian and Laplace mechanisms using an Edgeworth series.
Both techniques are very efficient in terms of memory and run time, even for very large datasets and across broad ranges of the mechanism parameters.
(3) We experimentally corroborate the accuracy of our bounds against a numerical ground truth, provide the first ReRo bounds for ImageNet-scale workflows and explain a finding by \cite{hayes2023bounding} regarding differences in ReRo bounds when DP-SGD parameters are varied at a fixed $(\varepsilon, \delta)$-value. 

\paragraph{Background}
We assume familiarity with the fundamentals of DP and omit a detailed introduction due to space constraints. 
In brief, we will focus on the global model of DP and the add/remove one adjacency relation between databases $D$ and $D'$.
An extension to replacement adjacency is straightforward.
We will denote the deterministic query function (e.g. a single step of SGD outputting a gradient containing sensitive information) by $q$ and its global sensitivity by $\Delta$ with an appropriate subscript to indicate the order of the norm it is measured in.
We will use $\mathcal{M}$ for an (additive noise) mechanism, i.e. the Laplace mechanism (LM), Gaussian mechanism (GM) or their Poisson-subsampled variants (SLM and SGM).
For details on subsampling, we refer to \cite{balle2018privacy}; in brief, to realise Poisson subsampling, each record in a database participates in the query with individual probability $p$.

In the \textbf{hypothesis testing interpretation of DP}, we presume that an adversary $\mathcal{A}$ who has complete knowledge of $D$, $D'$, $q$, and all specifications of $\mathcal{M}$ observes a mechanism output $y$ and must decide: $\mathcal{H}_0: y \sim \mathcal{M}(D)$ vs. $\mathcal{H}_1: y \sim \mathcal{M}(D')$.
$\mathcal{H}_0$ and $\mathcal{H}_1$ are called the null and alternative hypothesis, respectively.
We stress that the only unknown in the aforementioned hypothesis testing problem is the exact noise draw realised by $\mathcal{M}$.
The privacy guarantee of $\mathcal{M}$ thus expresses how difficult it is to distinguish between the distributions $\mathcal{M}(D)$ and $\mathcal{M}(D')$ as measured in terms of trade-off between the fundamental errors of hypothesis testing: the Type-I error $\alpha$ and the Type-II error $\beta$.
Since the aforementioned hypothesis testing problem is one between two simple hypotheses, $\mathcal{A}$ is endowed with the optimality properties furnished by the Neyman-Pearson (NP) lemma \cite{neyman1933ix}.
In other words, their test has the highest power $1-\beta$ at any given level $\alpha \in [0, 1]$.
$f$-DP \cite{dong2021gaussian} utilises a trade-off function $T: \alpha \mapsto \beta$ to express DP guarantees.
Concretely, let $\phi$ be a rejection rule for the aforementioned hypothesis testing problem.
Then, $T(\mathcal{M}(D), \mathcal{M}(D'))(\alpha) = \inf_{\phi} \lbrace \beta_{\phi} \mid \alpha_{\phi} \leq \alpha \rbrace$.
A mechanism is said to satisfy $f$-DP, if, for all $\alpha \in [0,1]$ and all adjacent $D, D'$ it holds that $T(\mathcal{M}(D), \mathcal{M}(D'))(\alpha) \geq f(\alpha)$, where $f$ is some \say{reference} trade-off function.
The $\inf_{\phi}$ means that, by definition, $f$-DP only considers the rejection rule with the highest power among all realisable rejection rules at the same level $\alpha$, which is consistent the optimality properties of $\mathcal{A}$.
For rejection rules with asymmetric trade-off functions (e.g. for sub-sampled mechanisms), one must also consider $T^{-1}=T(\mathcal{M}(D'), \mathcal{M}(D))$ and obtain the symmetrised/convexified curve $\mathrm{C}(T, T^{-1})$.
This is important as the DP guarantee must hold identically for the \say{add one} and the \say{remove one} adjacency relations.
A mechanism whose trade-off function is $\beta(\alpha)=1-\alpha$, i.e. the off-diagonal of the unit square, offers perfect privacy.
As a worst-case guarantee, $f$-DP thus additionally only considers the trade-off function which is farthest from this off-diagonal, corresponding to the pair of mechanism distributions exhibiting the greatest \textit{effect size}. 
This pair is called the \textit{dominating pair} of a mechanism \cite{zhu2022optimal}. 
For the GM, the dominating pair is $(\mathcal{N}(0, \sigma^2), \mathcal{N}(\Delta_2, \sigma^2))$ and for the LM it is $(\mathrm{Lap}(0, b), \mathrm{Lap}(\Delta_1, b))$.
For the SGM two pairs must be considered: $(\mathcal{N}(0, \sigma^2), (1-p)\mathcal{N}(0, \sigma^2)+p\mathcal{N}(\Delta_2, \sigma^2))$ and $((1-p)\mathcal{N}(\Delta_2, \sigma^2)+p\mathcal{N}(0, \sigma^2), \mathcal{N}(\Delta_2, \sigma^2))$; this transfers to the SLM by replacing the Gaussian by the Laplace density.

\textbf{ReRo} \cite{balle2022reconstructing} is an upper bound on the probability of a successful DR attack.
In this work, we will study ReRo under a pessimistic threat model which is very similar to that of DP:
$\mathcal{A}$ has access to all database records and executes a DR attack $R$ on a model $w$ outputting a reconstructed record $z^\ast \sim R(w)$.
The goal of $\mathcal{A}$ is to select the correct database record $z$ corresponding to $z^\ast$ (i.e. record matching).
Formally, let $\pi$ denote $\mathcal{A}$'s prior distribution (i.e. auxiliary information) and let $\rho$ be a reconstruction error function. 
Then, $\mathcal{M}$ satisfies $(\eta, \gamma)$-ReRo if, for any fixed $D$, it holds that $\mathbb{P}_{z\sim \pi, w \sim \mathcal{M}(D \cup \lbrace z \rbrace)}(\rho(z, R(w))\leq \eta) \leq \gamma$.
Note the difference to DP: ReRo is defined purely through the \say{add one} adjacency relation.
The authors of \cite{hayes2023bounding} directly show that mechanisms whose output distributions satisfy a bound on the so-called \say{blow-up function} $\mathcal{B}_{\kappa(\eta)}$ also satisfy ReRo.
Concretely, let $\mu$ and $\nu$ be $\mathcal{M}$'s dominating pair distributions for the \say{add one} adjacency relation and $E$ be a measurable event.
Then, $\mathcal{M}$ satisfies $(\eta, \gamma)$-ReRo with respect to a prior $\kappa(\eta)$ with $\gamma = \mathcal{B}_{\kappa(\eta)}(\mu, \nu) = \sup \lbrace \mathbb{P}_{\mu}(E) \mid \mathbb{P}_{\nu}(E) \leq \kappa(\eta) \rbrace$.
Throughout, we follow \cite{hayes2023bounding} and let $\rho=\boldsymbol{1}(z\neq z^\ast)$ (i.e. an exact match) and assign a uniform prior $\kappa(\eta)=1/n$, where $n$ can e.g. be the cardinality of the database, since $\mathcal{A}$ has an \textit{a priori} probability of $1/n$ to select the correct candidate record without observing $R(w)$, or some more pessimistic fixed prior, e.g. $1/10$.
Although general hypothesis testing theory is used in \cite{hayes2023bounding} to prove the ReRo bound for DP mechanisms, the authors do not directly use $f$-DP to bound ReRo and instead estimate $\gamma$ using MC (Algorithm 1 of \cite{hayes2023bounding}).
This strategy has the drawback of holding only at the limit as the number of MC samples approaches infinity and is impracticable for very large $n$ or very small $\kappa$.
Next, we will show that $\mathcal{B}_{\kappa(\eta)}(\mu, \nu)$ has a natural hypothesis testing interpretation, allowing us to circumvent the MC procedure and directly bound $\gamma$.

\section{ReRo bounds for DP mechanisms through hypothesis testing\hfill\mbox{}}
We begin by expressing $\mathcal{B}_{\kappa(\eta)}$ in terms of the hypothesis testing problem between $\mathcal{M}(D)$ and $\mathcal{M}(D')$.
Assume that $\mathcal{A}$ employs a rejection rule $\phi$ with power $1-\beta_{\phi}(\alpha)$ at a pre-selected level $\alpha$.
Consistent with the worst-case guarantee, we will only consider the rejection rule with the highest power among all realisable rejection rules and denote this \textit{supremum power} as $\mathcal{P}(\alpha)$.
We remark that we make no further specifications about the rejection rule.
Therefore, although we will consider the DP threat model which assumes an optimal $\phi$ using the likelihood ratio test statistic evaluated at the dominating pair, all results transfer to threat model relaxations, provided the realisable rejection rules and their corresponding test statistics can be specified.
(2) We formulate our results in terms of the test $\mathcal{H}_0:\mathcal{M}(D)$ vs. $\mathcal{H}_1:\mathcal{M}(D')$ because we only need to bound the \say{add one} adjacency relation to bound ReRo.
The upshot of this choice can be seen in Figure 1, panel \textbf{f}.
\begin{theorem}\label{thm1}
    If $\mathcal{M}$ upper-bounds the adversary's supremum power $\mathcal{P}(\alpha)$, then it also satisfies $(\eta, \mathcal{P}(\kappa(\eta))$-ReRo for a prior $\kappa$.
    In particular, if $\mathcal{M}$ satisfies $f$-DP, it also satisfies $(\eta, 1-f(\kappa(\eta)))$-ReRo and if it satisfies $(\varepsilon, \delta)$-DP, it also satisfies $(\eta, \min\lbrace \mathrm{e}^{\varepsilon}\kappa(\eta) + \delta, 1\rbrace)$-ReRo.
\end{theorem}
The special case of $(\varepsilon, 0)$-DP appeared previously in \cite{balle2018privacy, hayes2023bounding}. 
The theorem's main advantage is that it allows us to think about the relationship between DP and ReRo in terms of \textit{statistical power analysis}, for which robust tools and an extensive body of theory exist.
Moreover, it explains the finding by \cite{hayes2023bounding} that directly bounding ReRo using $\mathcal{B}_{\kappa(\eta)}(\mu, \nu)$ instead of taking a \say{detour} via Rényi DP results in a tighter bound: \textit{ReRo has a natural hypothesis testing interpretation, whereas Rényi DP does not} \cite{balle2020hypothesis}.
Furthermore, the theorem establishes that ReRo as a weaker guarantee than $f$-DP in the sense that $f$-DP bounds $\mathcal{A}$'s supremum power at all levels $\alpha \in [0,1]$, whereas ReRo is a bound on the supremum power at a single level $\alpha = \kappa(\eta)$.
Consequently, achieving ReRo is easier (i.e. requires less noise) than achieving $f$-DP.

In terms of concrete mechanisms, we obtain the following results:
\begin{corollary}
    Let $\mu_1 = \nicefrac{\Delta_1}{b}$ and
    \begin{equation}
        f_{\mathrm{Lap}}(\alpha, \mu_1) = 
        \begin{cases}
        1-\alpha\mathrm{e}^{\mu_1}, & \alpha < \nicefrac{\mathrm{e}^{-\mu_1}}{2} \\
        \nicefrac{\mathrm{e}^{-\mu_1}}{4\alpha}, & \nicefrac{\mathrm{e}^{-\mu_1}}{2} \leq \alpha \leq \nicefrac{1}{2} \\
        (1-\alpha)\mathrm{e}^{-\mu_1}, & \alpha > \nicefrac{1}{2}.
        \end{cases}
    \end{equation}
    Then, the LM satisfies $(\eta, \gamma)$-ReRo with $\gamma = 1-f_{\mathrm{Lap}}(\kappa(\eta), \mu_1)$.
\end{corollary}
\begin{corollary}
    Let $\mu_2 = \nicefrac{\Delta_2}{\sigma}$ and $f_{\mathrm{Gauss}}(\alpha, \mu_2) = \Phi(\Phi^{-1}(1-\alpha) - \mu_2)$, where $\Phi$ and $\Phi^{-1}$ are the cumulative distribution and quantile function of the standard normal distribution. Under $N$-fold homogeneous composition, the GM satisfies $(\eta, \gamma)$-ReRo with $\gamma = 1-f_{\mathrm{Gauss}}(\kappa(\eta), \sqrt{N}\mu_2)$.
    Under heterogeneous composition of mechanisms with $\mu_a, \mu_b, \dots$, we have $\gamma = 1-f_{\mathrm{Gauss}}(\kappa(\eta), \sqrt{\mu_a^2+\mu_b^2+\dots})$.
\end{corollary}
These two corollaries allow us to obtain an \textit{exact} bound on ReRo for the respective mechanisms. 
Unfortunately, the trade-off functions for the LM under composition and for the SLM and SGM are not available in closed form.
Three distinct options exist for evaluating these functions:
(1) Compute the trade-off functions numerically either through direct numerical integration or e.g. using the technique by \cite{doroshenko2022connect}.
This approach can be optimal in the sense that it can provide an exact bound up to numerical precision (or with a controlled error tolerance).
To obtain a valid ground truth, we use direct numerical integration by performing a grid discretisation over $G$ points and using an arbitrary-precision floating point library such as \cite{mpmath}. 
This technique is extremely time-consuming, as (for $N$ composition steps) it requires $G \cdot N$ numerical integrations (in neural network applications $N = \mathcal{O}(10^{4})$) and thus only serves as a gold standard.
An approach using the technique by \cite{doroshenko2022connect} can be found in the appendix.
(2) One can leverage an analytic (e.g. Edgeworth or saddle-point) \textit{finite sample approximation} to the trade-off function which can be computed in constant time for homogeneous composition.
Such approximations are a cornerstone of statistical power analysis \cite{cox1979theoretical}, and have been previously used for $(\varepsilon, \delta)$-DP accounting \cite{wang2022analytical, alghamdi2022saddle}.
For our experiments, we use an improved version of the technique proposed by \cite{zheng2020sharp}, i.e. a fourth order Edgeworth approximation, which has error $\mathcal{O}(N^{-2})$. 
(3) Asymptotically, the trade-off function of a (Poisson-)subsampled mechanism with sampling rate $p$ converges to $f_{\mathrm{Gauss}}(\alpha, \tilde{\mu})$ with $\tilde{\mu}= p\sqrt{N(\mathrm{e}^{\mu_2}-1)}$ when $p\sqrt{N}$ converges to a positive constant as the compositions $N\rightarrow\infty$ \cite{bu2020deep}.
This so-called \textit{CLT approximation} is essentially an order zero Edgeworth approximation and has an error of $\mathcal{O}(N^{-\nicefrac{1}{2}})$.
We note that, although the MC technique of \cite{hayes2023bounding} has a nominally even higher error rate of $\mathcal{O}((\kappa N)^{-\nicefrac{1}{2}})$, it performs better than the CLT approximation in practice because it is unbiased, whereas the CLT approximation presupposes that the approximated trade-off function is Gaussian, which leads to poor performance when its assumptions are violated (see experiments below and \cite{gopi2021numerical} for discussion).
Independent of the technique used to approximate the trade-off function, we can formulate the following results:
\begin{corollary}
    Let $\tilde{f}_{\mathrm{SLM}}(\alpha, \mu_1, N, p)$ denote the approximate trade-off function for the SLM with sampling rate $0<p\leq 1$ under $N$-fold composition using one of the approximation techniques above.
    Then, the SLM satisfies $(\eta, \gamma)$-ReRo with $\gamma \approx 1-\tilde{f}_{\mathrm{SLM}}(\kappa(\eta), \mu_1, N, p)$.
    Similarly, let $\tilde{f}_{\mathrm{SGM}}(\alpha, \mu_2, N, p)$ denote the approximate trade-off function for the SGM with sampling rate $0<p<1$.
    Then, the SGM satisfies $(\eta, \gamma)$-ReRo with $\gamma \approx 1-\tilde{f}_{\mathrm{SGM}}(\kappa(\eta), \mu_2, N, p)$.
\end{corollary} 

Note that for the SGM, when $p=1$, we revert to the GM and can use the closed-form bound from Corollary 2 (see Figure 1\textbf{c} below).
We remark for completeness that heterogeneous composition is also possible using the techniques above and that approximations are not necessarily valid upper bounds unless verified, e.g. using the technique by \cite{wang2023randomized}.
We omit a detailed discussion of these points due to space constraints.

\section{Experimental evaluation and conclusion}

\begin{figure}[htb]
  \centering
  \begin{subfigure}[b]{0.32\textwidth}
    \centering
    \includegraphics[width=\textwidth]{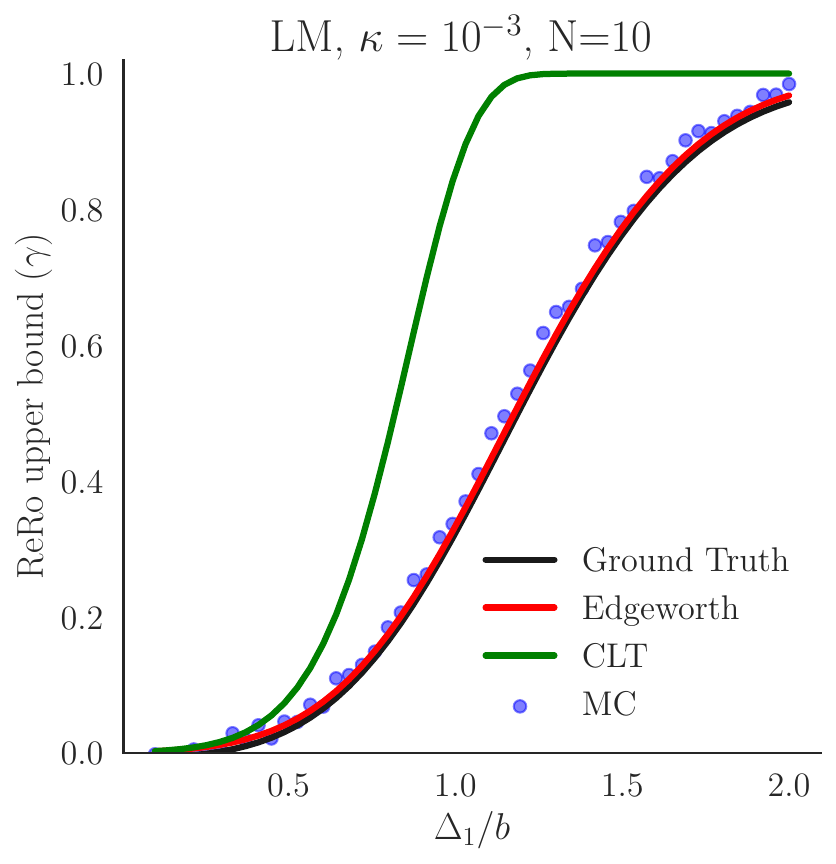}
    \caption{}
    \label{fig:subfig1}
  \end{subfigure}
  \begin{subfigure}[b]{0.32\textwidth}
    \centering
    \includegraphics[width=\textwidth]{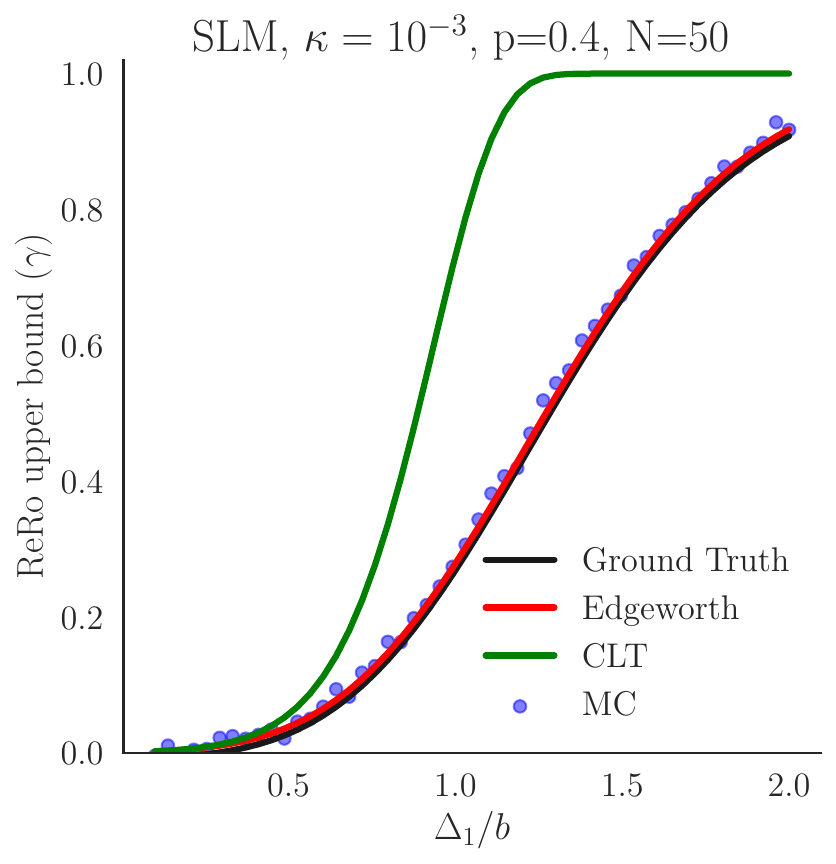}
    \caption{}
    \label{fig:subfig2}
  \end{subfigure}
  \begin{subfigure}[b]{0.32\textwidth}
    \centering
    \includegraphics[width=\textwidth]{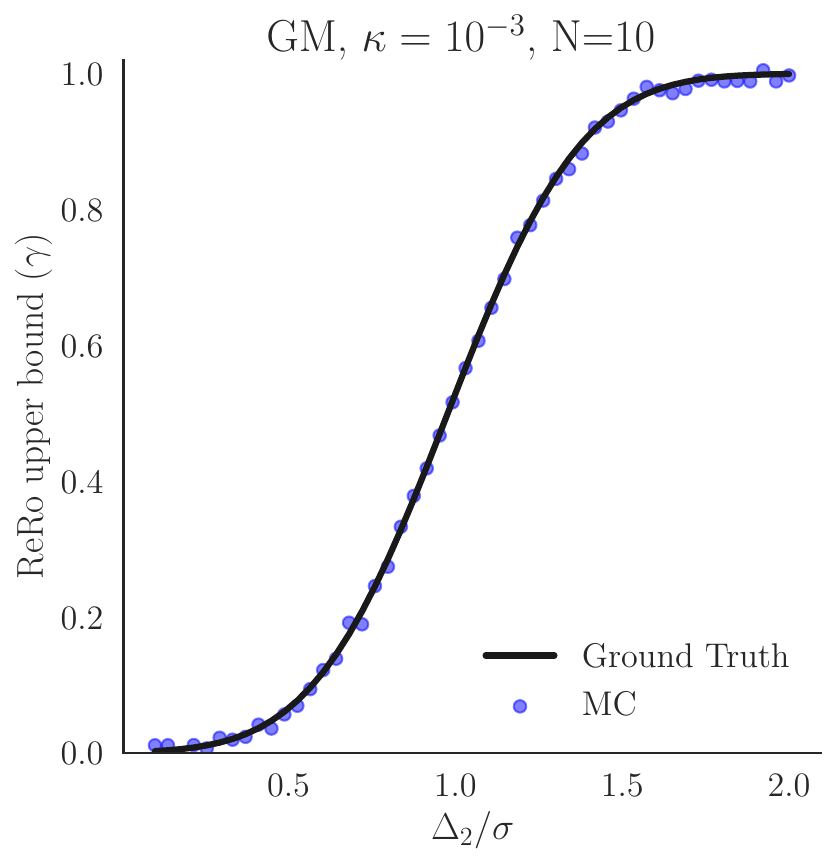}
    \caption{}
    \label{fig:subfig3}
  \end{subfigure}
  \begin{subfigure}[b]{0.32\textwidth}
    \centering
    \includegraphics[width=\textwidth]{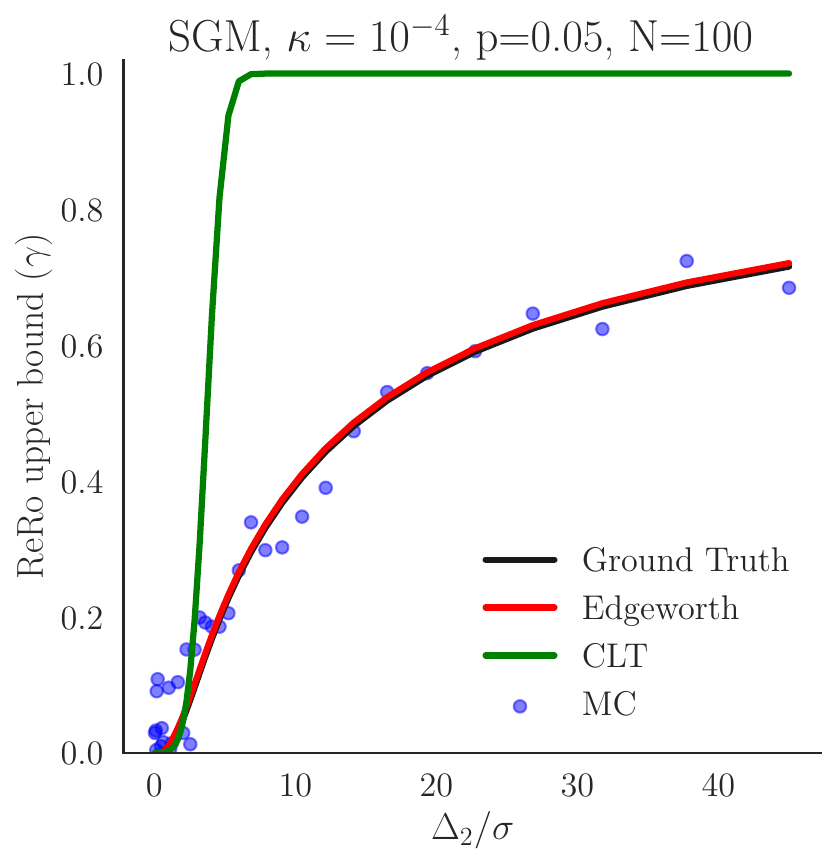}
    \caption{}
    \label{fig:subfig4}
  \end{subfigure}
  \begin{subfigure}[b]{0.32\textwidth}
    \centering
    \includegraphics[width=\textwidth]{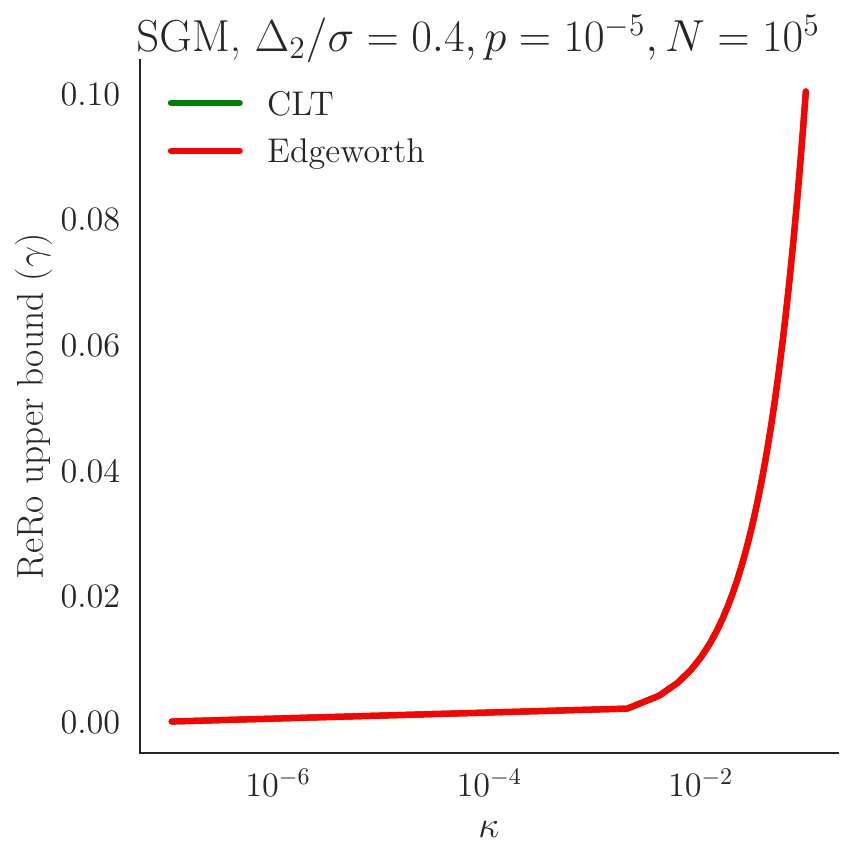}
    \caption{}
    \label{fig:subfig5}
  \end{subfigure}
  \begin{subfigure}[b]{0.32\textwidth}
    \centering
    \includegraphics[width=\textwidth]{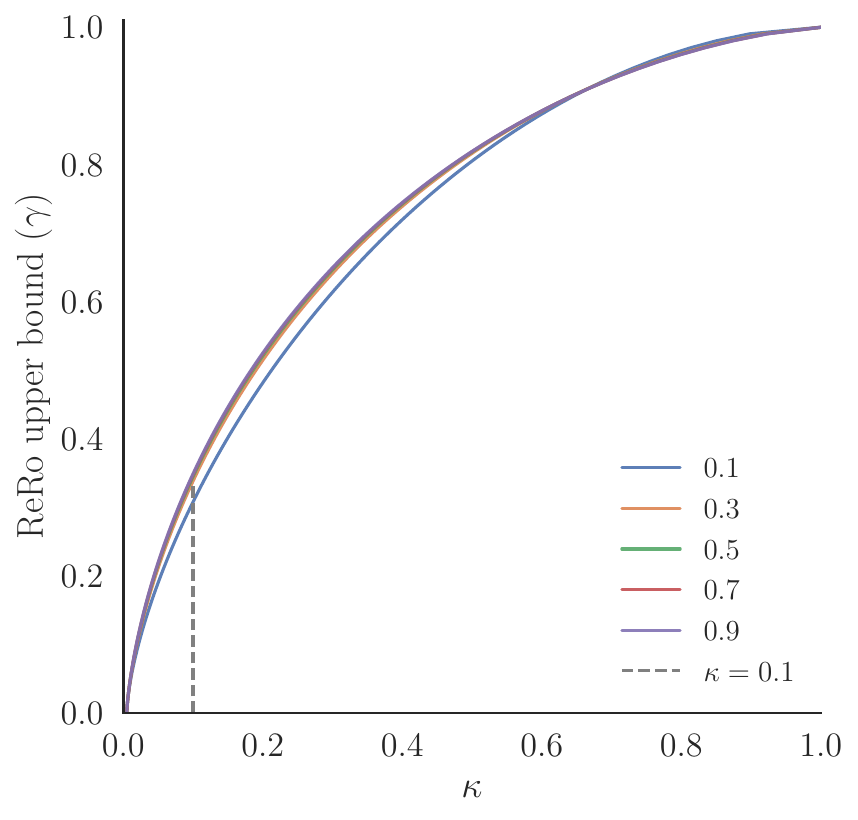}
    \caption{}
    \label{fig:subfig6}
  \end{subfigure}

  \caption{}
  \label{fig:mainfig}
\end{figure}
Figure \ref{fig:mainfig} compares the \textcolor{blue!70}{MC} estimate \cite{hayes2023bounding} of $\gamma$ ($10^{6}$ samples) at a fixed prior $\kappa$ to the asymptotic \textcolor{darkgreen}{CLT} approximation \cite{bu2020deep}, the fourth-order \textcolor{red}{Edgeworth} approximation and the \say{Ground Truth} computed by numerical integration (\textbf{a},\textbf{b},\textbf{d}) or in closed form (\textbf{c}).
$\gamma$ is plotted against the effect size ($\nicefrac{\Delta_1}{b}$ or $\nicefrac{\Delta_2}{\sigma}$), corresponding to increasing privacy loss: \textbf{a}: $\varepsilon_{\max}=20$, \textbf{b}/\textbf{c}: $\varepsilon_{\max}=100$, \textbf{d}: $\varepsilon_{\max}=\mathcal{O}(10^{8})$ at $\delta=10^{-6}$ for \textbf{c}/\textbf{d}.
Observe that in panel \textbf{d}, the MC algorithm of \cite{hayes2023bounding} already has too high variance to provide an accurate estimate of $\gamma$.
This means that the analysis of ImageNet-sized datasets where the values of $\kappa$ and $p$ are very low and the number of steps $N$ is very high is infeasible using MC (or the Ground Truth) due to memory or time constraints.
In contrast, estimating $\gamma$ using the Edgeworth approximation yields excellent precision at a constant memory consumption and run time of only about $1.5$s, exactly matching the Ground Truth.
Panel \textbf{e} shows $\gamma$ as a function of $\kappa$ for a very low $p$ and a very large $N$, similar to the hyperparameters used by \cite{de2022unlocking} when training ImageNet from scratch.
Even at $\kappa=10^{-7}$, our presented techniques allow for estimating $\gamma$, and the CLT approximation matches the Edgeworth approximation very well.
Further examples of CIFAR-10 and ImageNet workflows are shown in the Appendix.
Panel \textbf{f} explains the observation by \cite{hayes2023bounding}, where, at a constant $(\varepsilon, \delta)$, the authors find that different sampling rates $p$ lead to different values of $\gamma$.
The crux of this finding is that the authors of \cite{hayes2023bounding} choose mechanism parameter combinations which result in the same privacy guarantee in terms of a \textit{single} $(\varepsilon, \delta)$-pair (recall that SGMs are only identical if they correspond in \textit{all possible} $(\varepsilon, \delta)$-pairs). 
Thus, mechanisms with different $p$ are fundamentally distinct and thus lead to different $\gamma$ values across the range of $\kappa$.
In particular, the trade-off function (and thus the ReRo bound function) is increasingly asymmetric at low values of $p$.
As seen in the figure, for $\kappa=0.1$ (used by \cite{hayes2023bounding}), $\gamma$ is lower at $p=0.1$ (blue) compared to $0.9$ (lavender), matching Figure 6 of \cite{hayes2023bounding}.
A detailed discussion on this topic can be found in the Appendix.

\paragraph{Conclusion}
In this work, we expanded on the connection between ReRo and DP by leveraging hypothesis testing theory and techniques from statistical power estimation.
This allowed us to formulate refined ReRo bounds for relevant DP mechanisms and propose techniques to estimate them with high precision across a broad range of use-cases.  
Our results can thus help ML practitioners to evaluate the vulnerability of sensitive data processing systems against data reconstruction attacks, thereby increasing user trust.
In future work, we intend to assess ReRo bound tightness for large vision and language models/datasets, provide ReRo bounds in the shuffle model of DP and for individual privacy accounting schemes, expand our analysis to non-uniform priors other reconstruction error functions and heterogeneous compositions.

\printbibliography
\newpage

\section{Appendix}
\subsection{Proofs}
\paragraph{Proof of Theorem 1}
\begin{proof}
    Let $y$ be a mechanism output, $\mu, \nu$ be the dominating pair distributions of $\mathcal{M}$ and $\kappa(\eta) \in [0,1]$ be a prior.
    Since $E$ is an arbitrary measurable event, we can fix $E$ to be the event of rejecting $\mathcal{H}_0$ (this mirrors the event definitions in Corollary 3 of \cite{hayes2023bounding} and standard hypothesis testing theory). 
    Moreover, let $\phi$ be a rejection rule for $\mathcal{H}_0: y \sim \nu$ and $\mathcal{H}_1: y \sim \mu$.
    This is without loss of generality since $f$ can always be considered (or made) symmetric, and thus the following statements also hold when the role of the hypotheses is exchanged, although this is not required to bound ReRo, which only considers the \say{add one} adjacency relation.

    From the definition of ReRo, $\gamma = \mathcal{B}_{\kappa(\eta)}(\mu, \nu) = \sup \lbrace \mathbb{P}_{\mu}(E) \mid \mathbb{P}_{\nu}(E) \leq \kappa(\eta) \rbrace$.

    From our assumption above, $\mathbb{P}_{\mu}(E)=1-\beta_{\phi}$ (correctly reject $\mathcal{H}_0$ given $\mathcal{H}_1$) and $\mathbb{P}_{\nu}(E) = \alpha_{\phi}$ (wrongly reject $\mathcal{H}_0$ given $\mathcal{H}_0$).
    
    Substituting, we obtain $\gamma = \sup \lbrace 1-\beta_{\phi} | \alpha_{\phi} \leq \kappa(\eta) \rbrace$.
    In other words, $\gamma$ exactly corresponds to the supremum power of $\phi$ given a pre-selected bound on Type-I error rate, i.e. $\gamma = \mathcal{P}(\alpha)$, and thus a bound on $\gamma$ is implied by a bound on $\mathcal{P}(\alpha)$ with $\alpha=\kappa(\eta)$.
    
    To prove the ReRo bound implied by $f$-DP, we consider the definition of the trade-off function: $f(\kappa(\eta)): \inf \lbrace \beta_{\phi} | \alpha_{\phi} \leq \kappa(\eta))$.
    Since $f$ is convex, continuous and non-increasing on the unit square, $1-f(\kappa(\eta))=\sup \lbrace 1-\beta_{\phi} | \alpha_{\phi} \leq \kappa(\eta)) = \mathcal{P}(\alpha) = \gamma$.
    We note that the reverse does not hold in general: bounding ReRo through a bound on $\gamma$ implies a bound on $\mathcal{P}(\alpha)$ for a specific level $\alpha$, whereas $f$-DP implies a bound on $\mathcal{P}(\alpha)$ at \textit{all} levels $\alpha \in [0,1]$.
    
    To prove the the ReRo bound implied by $(\varepsilon, \delta)$-DP, we leverage a result by \cite{wasserman2010statistical}, who show that, if a mechanism satisfies $(\varepsilon, \delta)$-DP, it imposes a bound on the power $1-\beta$ at a level $\alpha$ of the optimal hypothesis test $\phi$ such that $1-\beta_{\phi}(\alpha_{\phi}) \leq \mathrm{e}^{\varepsilon}\alpha_{\phi} + \delta$, i.e. $\mathcal{P}(\alpha) = \mathrm{e}^{\varepsilon}\alpha_{\phi} + \delta$.
    Finally, we substitute $\kappa(\eta)$ as the desired level $\alpha_{\phi}$ and take the $\min$ since $\gamma$ is a probability, from which the claim follows. 
\end{proof}

\begin{remark}
Algorithm 1 of \cite{hayes2023bounding} essentially computes an MC estimate of the complementary trade-off function for $\mathcal{N}(0, \sigma^2)$ vs. $(1-p)\mathcal{N}(0, \sigma^2)+p\mathcal{N}(\Delta_2, \sigma^2)$.
The sampling inefficiency and high variance at small values of $\kappa$ is due to the fact that the algorithm draws $S$ MC samples but discards all but $\lceil \kappa \cdot S \rceil$ of them.
This percolates to extreme parameter regimes such as the ones discussed above, necessitating orders of magnitude more samples to be drawn to correctly estimate the bound, which eventually becomes infeasible due to memory constraints.
\end{remark}

\begin{remark}
In terms of the distributions of the likelihood ratio test statistics under $\mathcal{H}_0$ and $\mathcal{H}_1$, constructing $\mathcal{P}(\alpha)$ corresponds to the following steps: 
Per the Neyman-Pearson lemma, the optimal test $\phi$ is realised by thresholding the likelihood ratio test statistics.
Let $c$ be the critical value for rejecting $\mathcal{H}_0$.
Then, (1) determine the value of $c$ for which $\alpha_{\phi}(c)<\kappa(\eta)$ by computing the quantile function of the test statistic under $\mathcal{H}_0$ at $1-\kappa(\eta)$ and
(2) compute the value of the cumulative distribution function of the test statistic under $\mathcal{H}_1$ evaluated at $c$.
The likelihood ratios under $\mathcal{H}_0$ and $\mathcal{H}_1$ are also called the \textit{privacy loss random variables} in DP.
The equivalence between the privacy loss random variables and the test statistics from which the trade-off function $f$ is computed represents the intuitive link between $f$-DP, $(\varepsilon, \delta)$-DP and ReRo and reinforces the pivotal role of the privacy loss random variable.
The Edgeworth approximation utilises the cumulant generating functions of the likelihood ratio test statistics computed numerically, followed by a series approximation combined with a numerical inverse for the quantile function.
The CLT approximation is equivalent to an Edgeworth approximation of order zero, rendering it quite inflexible, which explains its poor performance when its assumptions are violated.
\end{remark}

\paragraph{Proof of Corollaries 1 and 2}
\begin{proof}
    The claims of both corollaries follow directly from the closed-form expressions of the trade-off functions of the LM and the GM. 
    The derivations of the trade-off functions themselves can be found e.g. in \cite{dong2021gaussian}.
\end{proof}

\paragraph{Proof of Corollary 3}
\begin{proof}
    The claims follow from the ReRo bound implied by $f$-DP proven in Theorem \ref{thm1}.
    We remark that since we are dealing with trade-off function approximations, minimising the approximation error is crucial for obtaining an exact bound on $\gamma$. 
\end{proof}

\subsection{Supplementary Figure}
The following figure illustrates further scenarios in which the Edgeworth and CLT approximation yield excellent results, whereas the MC technique of \cite{hayes2023bounding} would not be usable due to an impracticably high number of MC samples required to obtain an accurate estimate.
Moreover, in these scenarios, the numerical ground truth would take on the order of weeks to compute and is thus unavailable.
In contrast, the \textcolor{red}{Edgeworth} and \textcolor{blue!70}{CLT} approximations are computable in constant time.
Moreover, the assumptions of the CLT approximation \say{kick in} for these parameter values and thus the two methods yield identical results.
The top figure row shows ReRo bounds for CIFAR-10-style workflows with hyperparameters taken from Table 13 of \cite{de2022unlocking} (left) and an even smaller sampling rate (right), whereas the bottom row shows ImageNet-style workflows with the hyperparameters from Table 15 of \cite{de2022unlocking} (left) and an even smaller batch size (right).
The bottom right panel is identical to Figure 1, panel \textbf{e} in the main manuscript.
For all panels, $\kappa \in [10^{-7}, 10^{-1}]$.

\begin{figure}[h]
    \centering
    \includegraphics[width=0.9\textwidth]{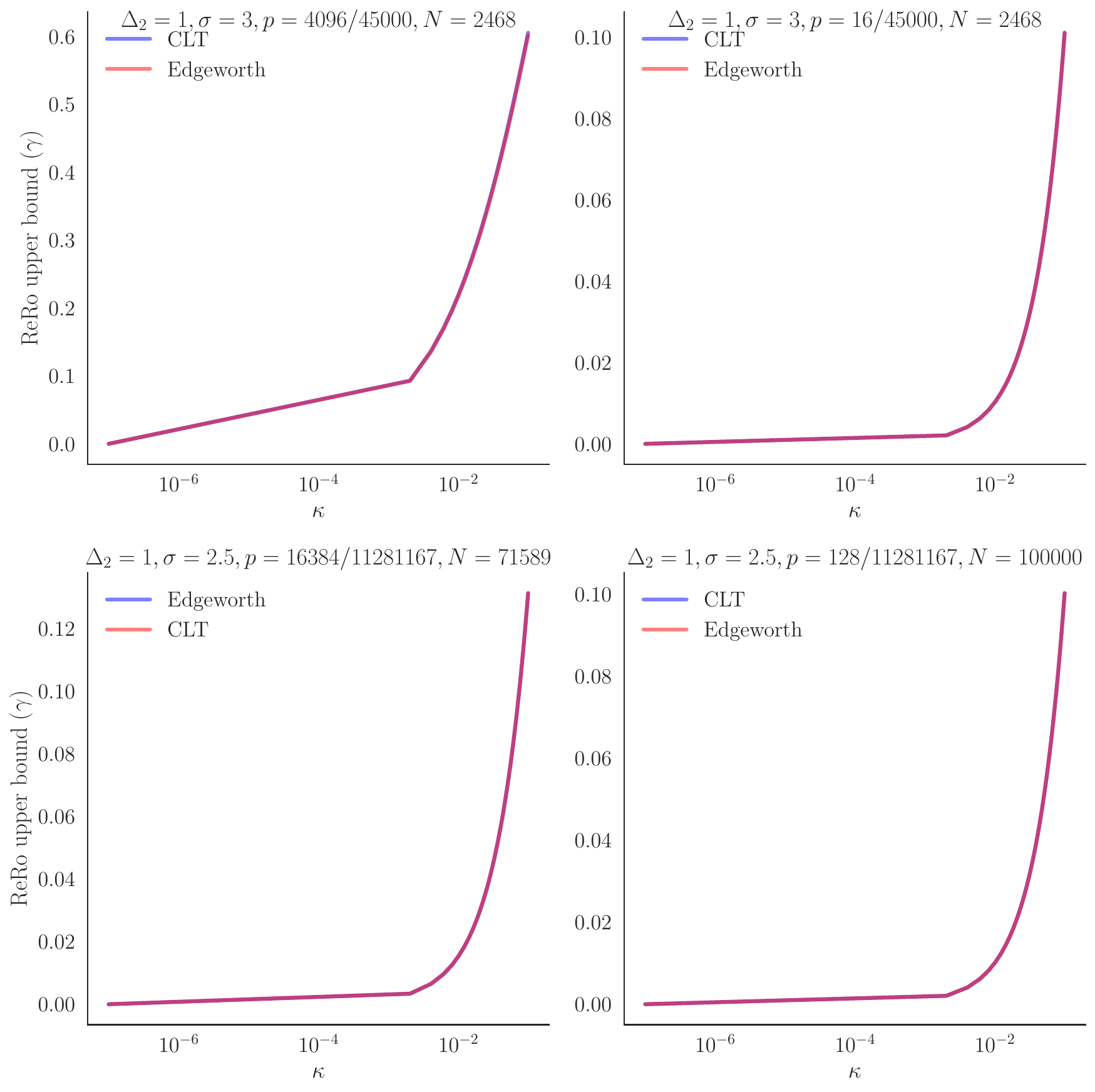}
    \caption{Bounding ReRo in CIFAR-10 and ImageNet-style workflows.}
\end{figure}

\subsection{Experimental details}

\paragraph{Conversion to $(\varepsilon, \delta)$-DP}
Conversions to $(\varepsilon, \delta)$-DP were performed as follows:
\begin{itemize}
    \item For the LM, following the simple composition theorem: $\varepsilon = \nicefrac{N\Delta_1}{b}$.
    \item For the SLM, following \cite{balle2018privacy}: $\varepsilon = \log(1 + p(\mathrm{e}^{\nicefrac{N\Delta_1}{b}}-1)$.
    \item For the GM, following \cite{dong2021gaussian}: Compute $\delta(\varepsilon) = \Phi\left(-\nicefrac{\sigma\varepsilon}{\Delta_2} + \nicefrac{\Delta_2}{2\sigma}\right) - \mathrm{e}^{\varepsilon}\Phi\left(-\nicefrac{\sigma\varepsilon}{\Delta_2} - \nicefrac{\Delta_2}{2\sigma}\right)$, then solve for $\varepsilon$ at a given $\delta$ numerically.
    \item For the SGM, following \cite{zhu2022optimal}: Compute the (symmetrised) trade-off function, then compute the convex conjugate numerically and solve for $\varepsilon$ at a given $\delta$.
\end{itemize}    

\paragraph{Details of numerical techniques}
The numerical Ground Truth was evaluated by using the technique proposed in Section 5.1 of \cite{zheng2020sharp} with $G=1\;000$ grid points and using $25$ digits of numerical precision in \cite{mpmath} (for reference, a $64$-bit floating point value provides $\approx 15$ digits of precision).
We recall that this technique requires one numerical integral per step $N$ and grid point, rendering it extremely time consuming and impracticable for any use beyond establishing a gold standard.
The fourth-order Edgeworth approximation was computed as previously described (see Section 3.1 of \cite{zheng2020sharp}).
However, we expanded the Edgeworth series up to order four as described in the main manuscript.
Moreover, the original work \cite{zheng2020sharp} only approximates the trade-off function for only one of the two dominating pairs of the SGM $(\mathcal{N}(0, \sigma^2), (1-p)\mathcal{N}(0, \sigma^2)+p\mathcal{N}(\Delta_2, \sigma^2))$.
Whenever required (e.g. for conversions to $(\varepsilon, \delta)$-DP or for Figure 1, panel \textbf{e}), we also instantiated the trade-off function for the other dominating pair $((1-p)\mathcal{N}(\Delta_2, \sigma^2)+p\mathcal{N}(0, \sigma^2), \mathcal{N}(\Delta_2, \sigma^2))$ and obtained the symmetrisation/convexification of the two trade-off functions, in line with the assumption that the trade-off function is symmetric.
Monte Carlo (MC) estimation of $\gamma$ was performed according to Algorithm 1 of \cite{hayes2023bounding}.
All MC experiments were performed with $S=1\, 000 \, 000$ samples.
We used multi-core sampling with 16 concurrent processes on a single 2019 Apple MacBook Pro with an 8 core Intel i9 CPU and 64 GB of memory.
The CLT and Edgeworth approximations have constant run time, the latter provided the composition is homogeneous (i.e. the effect size is constant over all $N$). 
In terms of memory usage, the MC algorithm allocates an array of size $S \cdot N$, where $S$ is the number of MC samples and $N$ is the number of SGM steps.
The numerical Ground Truth, Edgeworth and CLT approximations require constant memory.

\subsection{Discussion of ReRo bound sensitivity to subsampling probability}

\begin{figure}[h]
    \centering
    \includegraphics[width=0.9\textwidth]{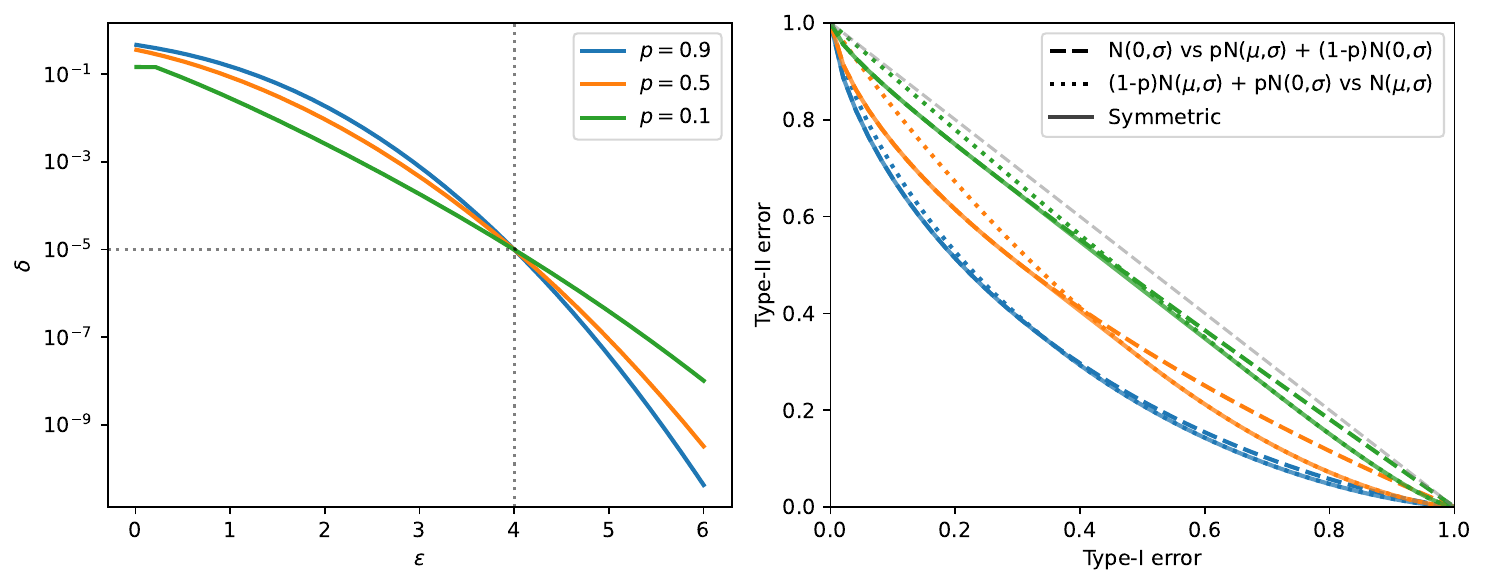}
    \caption{Comparing mechanisms at $p=0.1, 0.5, 0.9$, where we find $\sigma$ for $(\varepsilon=4, \delta=10^{-5})$-DP with the number of steps set to $N=1$. The mechanisms only intersect at $(\varepsilon=4, \delta=10^{-5})$-DP, meaning they are distinct and thus also result in distinct trade-off curves. The add and remove relations are also distinct, giving asymmetric trade-off functions.}
    \label{fig:no_clt}
\end{figure}

In \cite{hayes2023bounding}, the authors found that the ReRo upper bound is dependent on the subsampling probability, $p$.
They showed this by fixing the number of steps in DP-SGD and the gradient cliping norm, and finding a $\sigma$ that would give a fixed $(\varepsilon, \delta)$-DP guaranteee across different subsampling rates.
In \cite{hayes2023bounding}, authors chose a small number of steps ($100$) for this experiment, due to the computational overhead of their MC estimation method. 
For this relatively small number of composition, the CLT assumption for Gaussian DP is not yet fully in effect, meaning the mechanisms authors selected at different values of $p$ are not identical, they only intersect for a specific choice of $\varepsilon$ and $\delta$. 
We plot this in \Cref{fig:no_clt}, and show the corresponding trade-off curves for the \say{add one} and \say{remove one} adjacency relations along with their symmetrised version (see Definition F.1 in \cite{dong2021gaussian}).
When the CLT does not apply, this comparison is across fundamentally distinct mechanisms with different trade-off curves, and so the upper bounds for ReRo are different. 

\begin{figure}[h]
    \centering
    \includegraphics[width=0.9\textwidth]{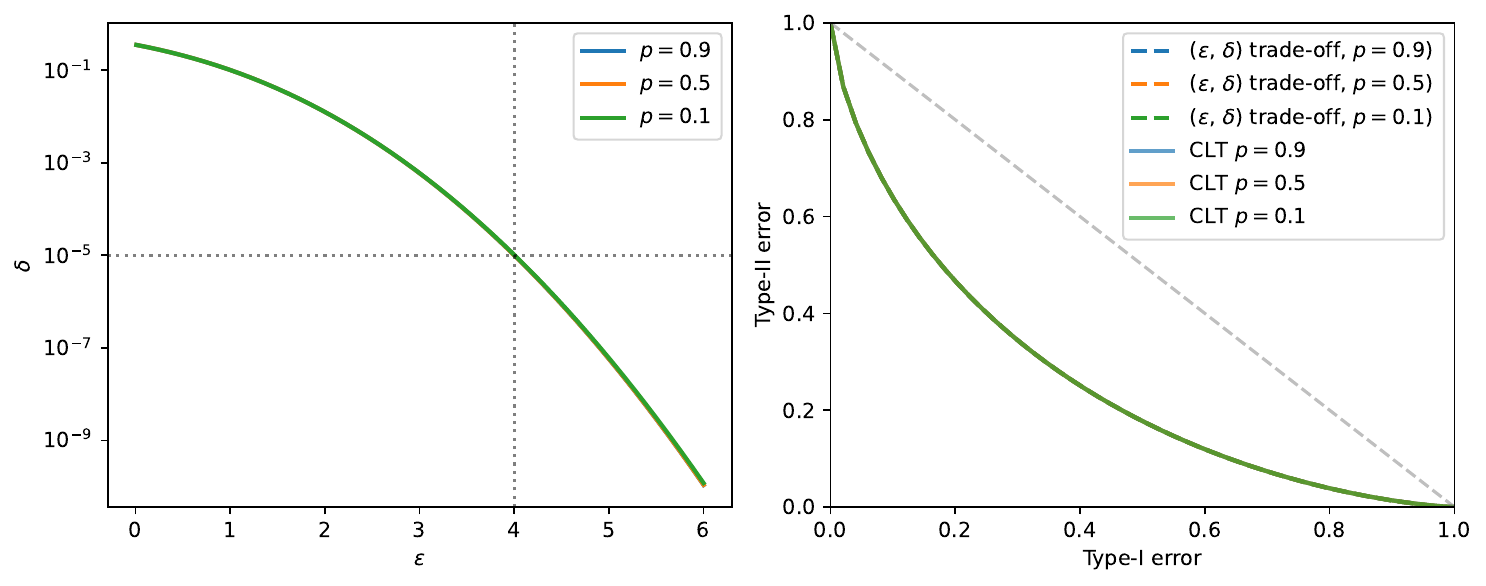}
    \caption{Comparing mechanisms at $p=0.1, 0.5, 0.9$, where we find $\sigma$ for $(\varepsilon=4, \delta=10^{-5})$-DP with the number of steps set to $N=10,000$. All three mechanisms are identical and have the same trade-off curves. In the right figure, we plot the trade-off curves under the assumption that the CLT holds~\cite{bu2020deep}, and when converting to a trade-off curve using the $(\varepsilon, \delta)$ trade-off function (Eq. 5 in \cite{dong2021gaussian}) with numerical accounting. Note there are three (indistinguishable) curves plotted in the left figure and six in the right figure.}
    \label{fig:clt}
\vspace{128in}
\end{figure}

We compare different values of subsampling probabilities when the CLT is assumed to hold (number of steps $N=10,000$).
From \Cref{fig:clt}, the three mechanisms intersect at identical $(\varepsilon, \delta)$ pairs, as so they are identical mechanisms.
In the right figure, we plot the trade-off curves under the assumption that the CLT is valid~\footnote{We use $
\tilde{\mu} = p \sqrt{N \left( \mathrm{e}^{1/\sigma^2} -1 \right)}$,
so the collection of all $\sigma$ to obtain the same $\tilde{\mu}$ can be found through:
$\sigma = \frac{1}{\sqrt{\log{\left(1 + \frac{\tilde{\mu}^{2}}{N p^{2}} \right)}}}$, see \cite{bu2020deep} for details.}. 
For each mechanism, we also numerically compute its privacy profile using~\cite{doroshenko2022connect} and convert to a trade-off curve using the $(\varepsilon, \delta)$ trade-off function (Eq. 5 in \cite{dong2021gaussian}). 
These all coincide perfectly.
When the CLT holds, the mechanisms are identical, the trade-off curves are independent of $p$, and since the curves are symmetric, the \say{add one} and the \say{remove one} curves are one and the same.

\end{document}